\documentclass[aps,prl,twocolumn,preprintnumbers,groupedaddress,superscriptaddress,floatfix,tightenlines,reprint,nofootinbib]{revtex4-1}
\usepackage{mathrsfs,natbib}
\usepackage{graphics,epsfig,color,graphicx}
\usepackage{verbatim}
\usepackage{bm}
\usepackage{enumerate}

\usepackage{graphicx,epsf,amssymb,bbm,amsbsy,amsfonts,amssymb,amsmath}
\usepackage{hyperref}

\newcommand{\eq}{\begin{equation}}
\newcommand{\eqe}{\end{equation}}

\hypersetup{
    colorlinks=true,       
    linkcolor=red,          
    citecolor=blue,        
    filecolor=magenta,      
    urlcolor=blue           
}

\def\OMIT#1{{}}

\newcommand{\beq}{\begin{equation}}
\newcommand{\eeq}{\end{equation}}
\newcommand{\beqa}{\begin{eqnarray}}
\newcommand{\eeqa}{\end{eqnarray}}

\newcommand{\Choose}[2]{{\begin{pmatrix} {#1} \\ {#2} \end{pmatrix}}}

\input xy
\xyoption{all}

\begin{document}

\title{Effective-range corrections to the ground-state energy \\ of the weakly-interacting Bose gas in two dimensions}

\preprint{NT-UW-17-16}

\author{\bf Silas R.~Beane}
\affiliation{Department of Physics, University of Washington
Seattle, WA 98195}

\begin{abstract}
  \noindent Nonuniversal effects due to leading effective-range
  corrections are computed for the ground-state energy of the
  weakly-coupled repulsive Bose gas in two spatial dimensions. Using
  an effective field theory of contact interactions, these corrections
  are computed first by considering fluctuations around the mean-field
  free energy of a system of interacting bosons. This result is then
  confirmed by an exact calculation in which the energy of a finite
  number of bosons interacting in a square with periodic boundary
  conditions is computed and the thermodynamic limit is explicitly
  taken.
\end{abstract}
\maketitle

{\bf Introduction.} Novel experimental techniques involving trapped
ultracold atoms allow the investigation of systems with reduced
dimensionality~\cite{RevModPhys.78.1111,RevModPhys.80.885,PhysRevLett.107.130401}.
In particular, experimental studies of dilute atomic gases in two
spatial dimensions are able to resolve elements of the
equation-of-state, including beyond mean-field
effects~\cite{Hadzibabic,PhysRevLett.102.170401,PhysRevLett.107.130401,Desbuquois}. This
of course motivates the theoretical study of atomic systems, with
their many types of interaction, as the spatial dimensionality is
altered. A valuable tool in this context that has enabled
model-independent descriptions of both bosonic and fermionic gases is
effective field theory (EFT) (for a relevant review see
Ref.~\cite{Kaplan:2005es}), which provides a means of systematically
improving quantum mechanical descriptions of fundamental properties of
atomic gases. In particular, EFT facilitates the calculation of
nonuniversal modifications, like many-body forces and shape-parameter
corrections, to the atomic
equation-of-state~\cite{RevModPhys.76.599,Braaten:2004rn}.  This
letter focuses on the calculation of effective-range (ER) corrections
to the ground-state energy of the weakly-interacting Bose gas in two
spatial dimensions. The universal effects in the ground-state energy
are known to three non-trivial orders in the weak coupling
expansion~\cite{PhysRevA.3.1067,Popov1972,PhysRevB.37.4936,Lieb2001,PhysRevE.64.027105,PhysRevLett.102.180404,Andersen2002}. In
addition there have been several studies of nonuniversal
effects~\cite{PhysRevA.81.013612,PhysRevLett.118.130402}.

Using the EFT as a starting point, the leading nonuniversal
corrections to the equation-of-state --due to ER corrections-- are
computed from the leading fluctuations about the mean-field result.
This calculation is a simple extension of the method used in three
spatial dimensions to two spatial
dimensions~\cite{Braaten:2000eh,RevModPhys.76.599}. However, the
two-dimensional derivation is somewhat more involved because of issues
related to renormalization, as will be seen below.  Remarkably, unlike
the case of three spatial dimensions, the two-dimensional ground state
energy can also be obtained exactly, without resorting to any
mean-field arguments~\cite{Beane:2010ny}. This is achieved by
calculating --perturbatively in the weak coupling constant-- the
energy of ${\bf N}$ bosons in a finite square with periodic boundary
conditions, and explicitly taking the thermodynamic limit. This
procedure relies on the tractability of the two-dimensional lattice
sums: they can be reduced to special functions with known
properties. Such a reduction of the three-dimensional lattice sums is
not known. It is found that the mean-field and exact results match
perfectly, as one would expect.

{\bf Effective field theory.}
The most general Lagrangian, constrained by Galilean invariance, parity and
time-reversal invariance, which describes Bosons interacting isotropically at low-energies
via an arbitrary finite-range potential is
\beqa
{\cal L}&=& {{\psi}^\dagger} \left( i\hbar \partial_t \; +\;  \frac{\hbar^2}{2M}{\bf\nabla}^2\; +\; \mu \right) \psi \\ \nonumber
&& - \frac{C_0}{4} ({{\psi}^\dagger} \psi)^2 \  -\  \frac{C_2}{8} {\bf\nabla}({{\psi}^\dagger } \psi){\bf\nabla}({{\psi}^\dagger } \psi) 
\ +\ \ldots.
\label{eq:1}
\eeqa 
Here three-body forces and higher-derivative operators have been
omitted. Throughout we use units with $\hbar =1$, and we keep the
boson mass, $M$, explicit. In $d=3$ spacetime dimensions, the mass
dimensions of the boson field and of the operator coefficients are
$[\psi]=1$, $[C_{2n}]=-1-2n$.

Using the two-body scattering conventions of Ref.~\cite{Beane:2010ny}, the ER expansion takes the form
\begin{eqnarray}
\cot\delta(p) \ = \ \frac{1}{\pi} \ln{\left(\frac{p^2}{\nu^2}\right)}  - \frac{1}{\alpha_{2}(\nu)} + \sigma_2\, p^2  +  {\mathcal O}( p^4 ) 
\label{eq:9}
\end{eqnarray}
where $\delta$ is the phase shift and
\begin{eqnarray}
\alpha_{2}(\nu )\ =\ \frac{M C_0(\nu )}{8}\ ; \qquad  \sigma_2\ =\ \frac{8C_2(\nu)}{M C_0^2(\nu )} 
\label{eq:9b}
\end{eqnarray}
are the ER parameters, written in terms of the renormalized EFT parameters that are defined using dimensional regularization with 
$\overline{MS}$~\cite{Beane:2010ny}. Here $\nu$ is the renormalization scale, $\alpha_2$ is a scale-dependent dimensionless coupling constant,
and $\sqrt{|\sigma_2|}$ is the effective range. Unlike the three-dimensional case, in $\overline{MS}$ in two spatial dimensions, all of the EFT
parameters are scale dependent. The leading beta function in the EFT is
\begin{eqnarray}
\nu\frac{d}{d\nu}C_0(\nu )\ = \ \frac{M}{4\pi}{C_0^2}(\nu )\ , 
\label{eq:10a}
\end{eqnarray}
which integrates to give the exact renormalization group (RG) evolution equation 
\begin{eqnarray}
\alpha_2(\nu )\ = \ \frac{\alpha_2(\nu')}{1- \frac{2}{\pi}\alpha_2(\nu' )\ln\left(\frac{\nu}{\nu'}\right)} \ .
\label{eq:rg1}
\end{eqnarray}

{\bf Free energy from mean-field fluctuations.} The technology for
computing the range corrections to the free energy of the
weakly-coupled Bose gas in the EFT in the case of three spatial
dimensions is well known~\cite{Braaten:2000eh,RevModPhys.76.599}. In
that case, in the $\overline{MS}$ scheme the EFT coefficients are
scale independent. The derivation is somewhat more subtle here due to the non-trivial
RG evolution in two spatial dimensions. The mean-field free energy and its
leading correction in the absence of ER corrections can
be expressed in $d$ spacetime dimensions with $\epsilon\equiv 3-d$ as
\begin{eqnarray}
\!\!\!\!\!\!{\cal F} = -\frac{\mu^2}{C_0} \; +\;  \frac{1}{4M}\left(\frac{\nu}{2} \right)^\epsilon\int \frac{d^{d-1}p}{(2\pi)^{d-1}} p\left( p^2 + 4\mu M \right)^{1/2}
\label{eq:Fbare}
\end{eqnarray}
where the first term is the mean-field result written in terms of the bare parameter $C_0$ and the second term is related to the sum of the zero-point energies of
the quasi-particles. With the renormalization scheme for $\alpha_2$ adopted in Ref.~\cite{Beane:2010ny} one easily obtains 
\begin{eqnarray}
{\cal F} = -\frac{\mu^2 M}{8 \alpha_2(\nu)} + \frac{\mu^2 M}{8 \pi}\bigg\lbrack \frac{1}{2}  + \ln\frac{\mu M}{\nu^2} \bigg\rbrack \ .
\label{eq:freenorange}
\end{eqnarray}
By Legendre transform one finds the mean-field and leading-loop
contributions to the ground-state energy
density~\cite{PhysRevA.3.1067,Popov1972,PhysRevB.37.4936,Lieb2001,PhysRevLett.102.180404,Andersen2002}
\begin{eqnarray}
{\cal E} \ =\ \frac{2\alpha_2(\nu)\rho^2}{M} \Bigg\lbrack\  1\ + \ 
\frac{\alpha_2(\nu)}{\pi}\Bigl( \frac{1}{2} + \ln\frac{4\alpha_2(\nu)\rho}{\nu^2}   \Bigr) \Bigg\rbrack.
\label{eq:Enorange}
\end{eqnarray}

The ER corrections can then be included by shifting the momenta under the square-root in Eq.~(\ref{eq:Fbare}) by
${\bf p}\rightarrow \Sigma_2^{1/2}{\bf p}$ where~\footnote{Note that while $\Sigma_2$ is renormalization scale dependent, this dependence
is subleading in $\alpha_2$.}
\begin{eqnarray}
\Sigma_2 \equiv 1\ +\ 2 M \frac{C_2(\nu )}{C_0(\nu )}\mu \ =\  1\ +\ 2 M \alpha_2(\nu)  \sigma_2 \mu \ .
\label{eq:shift}
\end{eqnarray}
The free energy with ER corrections is found to be
\begin{eqnarray}
{\cal F} = -\frac{\mu^2 M}{8 \alpha_2(\nu)} + \frac{\mu^2 M}{8 \pi \Sigma_2^{3/2}}\bigg\lbrack \frac{1}{2}  + \ln\frac{\mu M}{\nu^2} - \ln\Sigma_2 \bigg\rbrack \ .
\label{eq:freefull}
\end{eqnarray}
The shift in momentum requires a change in the renormalization scheme. The corresponding modification of the running of the coupling is:
\begin{eqnarray}
\alpha_2(\nu )\ = \ \frac{\alpha_2(\nu')}{1- \frac{2}{\pi}\alpha_2(\nu' )\Sigma_2^{-3/2}\ln\left(\frac{\nu}{\nu'}\right)} \ ,
\label{eq:rg1mod}
\end{eqnarray}
from which it is verified that Eq.~(\ref{eq:freefull}) is scale independent up to neglected universal ${\cal O}(\alpha_2)$ corrections.
We will see below in the exact calculation how this density-modified RG evolution is reconciled with the two-body evolution equation of Eq.~(\ref{eq:rg1}).

Expanding to leading order in the effective range gives~\footnote{It is not necessary to expand $\Sigma_2$ in Eq.~(\protect\ref{eq:freefull}). However,
there is little point in keeping the subleading terms as they are expected to be of the same size as shape parameter corrections, 
which have been neglected but are not generally expected to vanish.}
\begin{eqnarray}\label{eq:freeexpand}
{\cal F}& =& -\frac{\mu^2 M}{8 \alpha_2(\nu)} + \frac{\mu^2 M}{8 \pi}\bigg\lbrack \frac{1}{2}  + \ln\frac{\mu M}{\nu^2}\bigg\rbrack \\ \nonumber
&&\qquad\qquad  - \frac{3 \mu^3 M^2 \alpha_2(\nu) \sigma_2 }{8 \pi}\bigg\lbrack \frac{7}{6}  + \ln\frac{\mu M}{\nu^2}\bigg\rbrack \ .
\end{eqnarray}
This is the main new result in this letter. Performing the Legendre transform yields the ER contribution to the energy density
\begin{eqnarray}
{\cal E}^{\sigma_2}& =& -\frac{24\alpha_2(\nu)^4 \rho^3\sigma_2}{\pi M }\bigg\lbrack \frac{7}{6}  + \ln\frac{4\alpha_2(\nu)\rho}{\nu^2}\bigg\rbrack \ .
\label{eq:energysig2}
\end{eqnarray}
It is instructive to check that the scaling of this contribution is
sensible~\cite{Braaten:2000eh,RevModPhys.76.599,Braaten:1996rq}. The
momentum operator in the mean field scales as ${\bf
  p}\sim{\bf\nabla}\sim \sqrt{\alpha_2\rho}$ as is evident from
Eq.~(\ref{eq:Enorange}).  Clearly the $C_2$ operator does not
contribute at the level of the mean field and therefore ER corrections
must arise from a loop. Again from Eq.~(\ref{eq:Enorange}), each loop
gives a factor of $\alpha_2\ln\alpha_2$.  And of course
${\psi}^\dagger \psi$ scales as $\rho$ and the $C_2$ coefficient
scales as $\sigma_2\alpha_2^2$. We expect the ER corrections to scale
as an insertion of the $C_2$ operator times a loop which gives:
$(\sigma_2\alpha_2^2)(\sqrt{\alpha_2\rho})^2(\rho)^2(\alpha_2\ln\alpha_2)\sim
\alpha_2^4\ln\alpha_2\rho^3 \sigma_2$, as found above. The cubic
dependence on the density implies that the leading ER corrections
provide an effective three-body force, albeit one that is highly
suppressed at weak coupling. Below we will confirm
Eq.~(\ref{eq:energysig2}) by an exact calculation in which no mean
field is assumed~\footnote{The result found here for the leading ER corrections is in disagreement
with Ref.~\protect\cite{PhysRevLett.118.130402}, which finds an effective-range contribution that is
enhanced by a power of $\alpha_2$, in violation of the basic scaling arguments. Evidently, this error is due
to subtleties in regulating the divergent integral using dimensional regularization.}.

It may prove useful to express current knowledge of the free
energy in a more common notation. Using the definition of the
ER parameters in Ref.~\cite{0305-4470-17-3-020,Braaten:2004rn}, we identify
\begin{eqnarray}
\alpha_2(\nu)\ =\ \pi \left( \ln 4/\left( a^2\nu^2 e^{2\gamma}\right) \right)^{-1} \  , \ \sigma_2\ =\ \frac{r_e^2}{2\pi} \ ,
\label{eq:Braatconv}
\end{eqnarray}
where $a$ is the scattering length and $r_e$ is the effective range. Choosing the renormalization scale $\nu^2=\mu M$ so
that the logarithms in Eq.~(\ref{eq:freeexpand}) vanish and, following Ref.~\cite{PhysRevLett.102.180404}, defining the new coupling
\begin{eqnarray}
{\boldsymbol{\epsilon}}(\mu)^{-1}\equiv \pi \alpha_2(\sqrt{\mu M})^{-1}-1 \ ,
\label{eq:moracouple}
\end{eqnarray}
then gives current knowledge of the free energy of the weakly-coupled, two-dimensional Bose gas, including the new nonuniversal contribution
from range corrections:
\begin{eqnarray}
\!\!\!\!{\cal F}& =& -\frac{\mu^2 M}{8 \pi} \bigg\lbrack \frac{1}{{\boldsymbol{\epsilon}}}+\frac{1}{2} + {\boldsymbol{\epsilon}}\left( \frac{8 I}{\pi}+\frac{7}{4}\mu M r_e^2\right) + 
{\mathcal O}( {\boldsymbol{\epsilon}}^2 ) \bigg\rbrack 
\label{eq:freemora}
\end{eqnarray}
where the ${\mathcal O}( {\boldsymbol{\epsilon}} )$ universal
contribution calculated (numerically with $I\simeq 1.005$) in
Ref.~\cite{PhysRevLett.102.180404} has been included. It is shown in
this reference that Monte Carlo
simulations~\cite{PhysRevA.71.023605,PhysRevA.79.051602} are able to
resolve the ${\mathcal O}( {\boldsymbol{\epsilon}} )$ universal
contribution. It would be interesting to see whether the ER
contribution can be similarly detected in numerical simulations. This
may prove challenging as the range corrections to the energy are
suppressed by one power of the weak coupling as well as one power of
the density --which must be extremely dilute in order to achieve the
weak-coupling regime-- as compared to the known universal
effects~\cite{PhysRevLett.102.180404}.

{\bf Two-body energy in a square.}  We now proceed with an exact
calculation of the energy density.  In a finite square area ($L^2$)
with periodic boundary conditions, the energy levels for the two-boson
system determine the phase shift through the eigenvalue
equation~\cite{Fiebig:1994qi,Beane:2010ny}
\begin{eqnarray}
\cot\delta(p) \ = \ \frac{1}{\pi^2}  \Bigg\lbrack {\cal S}_2\,\left( \frac{p L}{2 \pi} \right) \ +\ 
2\pi\ln{\left(\frac{p L}{2 \pi} \right)} \Bigg\rbrack \ ,
\label{eq:24}
\end{eqnarray}
where
\begin{eqnarray}
{\cal S}_2\left( \eta \right)\ \equiv \  \sum^{\Lambda_n}_{{\bf n}\in\mathbb{Z}^2} \frac{1}{ {\bf n}^2 - \eta^2} \ -\ 2\pi \ln\Lambda_n \ .
\label{eq:25}
\end{eqnarray}
Unlike the case of three spatial dimensions, this integer sum is
tractable and indeed can be expressed in terms of the digamma
function~\cite{Beane:2010ny}. 

Neglecting shape-parameter corrections, the low-energy expansion, Eq.~(\ref{eq:9}), combined with
the eigenvalue equation, Eq.~(\ref{eq:24}), gives
\begin{eqnarray}
-\frac{1}{\alpha_2(\nu)} - \frac{2}{\pi} \ln{\left(\frac{\nu L}{2\pi}\right)}  + \sigma_2\, p^2 
 =  \frac{1}{\pi^2} {\cal S}_2\,\left( \frac{p L}{2 \pi} \right) \; .
\label{eq:26}
\end{eqnarray}
Using the RG evolution equation, Eq.~(\ref{eq:rg1}), it follows that
\begin{eqnarray}
-\frac{1}{\alpha_2}\ +\ \sigma_2\, p^2  \ = \ \frac{1}{\pi^2} {\cal S}_2\,\left( \frac{p L}{2 \pi} \right) \ ,
\label{eq:27a}
\end{eqnarray}
where $\alpha_2\equiv\alpha_2({2\pi}/{L})$. As the scale of the
coupling is fixed to $2\pi/L$, as the continuum limit is approached,
the repulsive theory is at weak coupling.
Hence when the two-body interaction is repulsive, the eigenvalue equation,
Eq.~(\ref{eq:27a}), allows a perturbative expansion of the energy
eigenvalues in the coupling $\alpha_2$.  In weak-coupling perturbation
theory, the ground-state energy is
\begin{eqnarray}
&&E_0\;+\;E_0^{\sigma_2} \ =\ \\ \nonumber && \frac{4\alpha_2}{M L^2} \Bigg\lbrack  1 -  \left(\frac{\alpha_2}{\pi^2}\right) {\cal P}_2 + 
 \left(\frac{\alpha_2}{\pi^2}\right)^2 \left( {\cal P}_2^2 - {\cal P}_4\right)  
 + {\cal O}(\alpha_2^3) \Bigg\rbrack  \\ \nonumber && + \frac{16\,\alpha_2^3\,\sigma_2}{M L^4}\Bigg\lbrack 1 - \left(\frac{\alpha_2}{\pi^2}\right) 3 {\cal P}_2 + {\cal O}(\alpha^2_2) \Bigg\rbrack \ ,
\label{eq:27PT}
\end{eqnarray}
where~\cite{Beane:2010ny}
\begin{eqnarray}
{\cal P}_2 &\equiv&  \sum^{\Lambda_n}_{{\bf n}\in\mathbb{Z}^2\neq 0} \frac{1}{ {\bf n}^2}  - 2\pi \ln\Lambda_n  =  4\pi \ln\left(e^{\frac{\gamma}{2}} \pi^{-\frac{1}{4}}\Gamma\left(\textstyle{\frac{3}{4}}\right)\right)\; ; \ \nonumber \\ 
{\cal P}_{2s} &\equiv& \sum^\infty_{{\bf n}\in\mathbb{Z}^2\neq 0} \frac{1}{ {(\bf n}^2)^s}  =  4 {\bf \zeta}(s){\bf \beta}(s) \quad ( s>1 )\ .
\label{eq:27PT2}
\end{eqnarray}
Here $\gamma$ is Euler's constant, $\Gamma(x)$ is the gamma function,  $\Lambda_n$ is an integer cutoff, and
\begin{eqnarray}
\!\!\!\!\!\!\!\!{\bf \zeta}(s)\ \equiv \ \sum_{m=0}^\infty \frac{1}{(m+1)^s} \ \ , \ \ {\bf \beta}(s)\ \equiv \ \sum_{m=0}^\infty \frac{(-1)^m}{(2m+1)^s} 
\label{eq:29}
\end{eqnarray}
are the Riemann zeta function and Dirichlet beta function, respectively.  

{\bf Many-body energy in a square.} 
The two-particle energy can be generalized to the ${\bf N}$-body system at
weak coupling using Rayleigh-Schr\"odinger perturbation theory~\cite{Beane:2007qr,Detmold:2008gh,Beane:2010ny} giving the ${\bf N}$-body ground-state energy
\begin{eqnarray}\label{eq:Nboson3}
&& E_0 + E^{\sigma_2}_0\ =\ \frac{4\,\alpha_2}{M L^2}{{\bf N} \choose 2}   \Bigg\lbrack\  1 -  \left(\frac{\alpha_2}{\pi^2}\right) {\cal P}_2\\ \nonumber 
&&\qquad + \left(\frac{\alpha_2}{\pi^2}\right)^2 \Bigl( {\cal P}_2^2 + (2{\bf N}-5) {\cal P}_4\Bigr)  
\ + {\cal O}(\alpha_2^3) \Bigg\rbrack \\ \nonumber
&& + \frac{16\,\alpha_2^3\,\sigma_2}{M L^4}{{\bf N} \choose 2} \Bigg\lbrack  1 +  \left(\frac{\alpha_2}{\pi^2}\right) 3({\bf N}-3) {\cal P}_2 +  {\cal O}(\alpha_2^2) \Bigg\rbrack 
\end{eqnarray}
where ${\tiny \Choose{n}{k}}$=$n!/(n-k)!/k!$.

{\bf Thermodynamic limit.}  
The goal in what follows is to take the thermodynamic limit of
Eq.~(\ref{eq:Nboson3}), where ${\bf N}$ and $L$ are taken to infinity
with the density, $\rho\equiv{\bf N}/L^2$, held fixed. This limit has
been taken explicitly in Ref.~\cite{Beane:2010ny} to obtain the
density expansion of the universal ground-state energy. Here we will
do the same to obtain the leading ER corrections.  Several issues
should be kept in mind. Firstly, naively, in the thermodynamic limit
only the subleading ER correction of ${\cal O}(\alpha_2^4)$ survives
the thermodynamic limit of $E^{\sigma_2}_0/{\bf N}$. While this is close to
the expected scaling, the result depends on the geometric constant
${\cal P}_2$. Clearly taking the thermodynamic limit must erase all
dependence on the geometry and thus must be independent of all of
the ${\cal P}_{2s}$ constants. Secondly, the coupling $\alpha_2$ is
evaluated at the far infrared scale $2\pi/L$, and therefore a change
of scale to a quantity which is fixed in the thermodynamic limit is
necessary.  It is straightforward to extend the ER contributions to
include the most-singular higher-order contributions giving
\begin{eqnarray}
&& E^{\sigma_2}_0\ =\ \frac{16\,\alpha_2^3\,\sigma_2}{M L^4}{{\bf N} \choose 2} \Bigg\lbrack\  1 +  \left(\frac{\alpha_2}{\pi^2}\right) 3({\bf N}-3) {\cal P}_2 \\ \nonumber
&& - \left(\frac{\alpha_2}{\pi^2}\right)^2\left( 10{\bf N}^2 {\cal P}_4 + \ldots\right)
+ \left(\frac{\alpha_2}{\pi^2}\right)^3\left( 35{\bf N}^3 {\cal P}_6 + \ldots\right)  +   {\cal O}(\alpha_2^4) \Bigg\rbrack 
\label{eq:range2}
\end{eqnarray}
where the dots above correspond to less-singular terms at that order in $\alpha_2$.
Now with $z\equiv{\bf N}\alpha_2/{\pi^2}$, we have
\begin{eqnarray}
\!\!\!\!\! E^{\sigma_2}_0 = \frac{16\,\alpha_2^3\,\sigma_2}{M L^4}{{\bf N} \choose 2} \Bigg\lbrack  1 +  3 z\left(1-\frac{3}{\bf N}\right) {\cal P}_2
 +  {\cal K}(z) \Bigg\rbrack
\label{eq:range3}
\end{eqnarray}
where
\begin{eqnarray}
{\cal K}(z)\; \equiv\; \sum_{n=2}^\infty(-1)^{n+1} {2n+1 \choose {n+1}}z^n{\cal P}_{2n} \ ,
\label{eq:range4}
\end{eqnarray}
and in Eq.~(\ref{eq:range3}) we have neglected subleading terms of ${\cal O}(\alpha_2^2)$ within the brackets.
Using an integral representation of the binomial coefficient and the representation, Eq.~(\ref{eq:27PT2}), of the
two-dimensional lattice sums, this function takes the form
\begin{eqnarray}
{\cal K}(z)\; =\; -\frac{16}{\pi} \int_0^\infty \frac{d\omega}{(1+\omega^2)^2}\sum_{n=2}^{\infty}{\bar z}^n \xi(n)\beta(n) \ ,
\label{eq:range5}
\end{eqnarray}
where ${\bar z}\equiv -4z/(1+\omega^2)$. Noting that
\begin{eqnarray}
&& \sum_{n=2}^{\infty}{\bar z}^n \xi(n)\beta(n)\; =\; \\ \nonumber
&&\qquad  {\bar z}\left( -\frac{\pi\gamma}{4}\ -\ 
\sum_{\ell=0}^\infty \frac{(-1)^\ell}{(2\ell +1)}\;\psi_0\left( 1 - \frac{\bar z}{(2\ell +1)}\right) \right) \ ,
\label{eq:app1d}
\end{eqnarray}
and using the asymptotic form of the digamma function for large argument gives
\begin{eqnarray}
\!\!\!\!\!\!{\cal K}(z) = -z\left( 3{\cal P}_{2}+ {\textstyle \frac{7}{2}}\pi  + 3\pi\ln z + \frac{1}{z}\right)  +  {\cal O}(1/z),
\label{eq:range5mod}
\end{eqnarray}
which clearly removes the geometric constant ${\cal P}_{2}$ from the energy, as it must.

In these expressions, the coupling $\alpha_2$ is evaluated at the
scale $2\pi/L$. Therefore, a change of scale is necessary in order to
take the thermodynamic limit. Say $\nu = 2\pi \sqrt{\lambda\rho}$,
where $\lambda$ is an arbitrary number. Now note that ${\cal K}(z)$
contains a contribution that scales as ${\bf N}\ln {\bf N}$ which will
not vanish in the thermodynamic limit if the coupling runs as in
Eq.~(\ref{eq:rg1}).  This singular piece is eliminated only if the
running of the coupling is modified so that with ${\alpha_2'}\equiv
\alpha_2(2\pi \sqrt{\lambda\rho})$, the RG evolution is
\begin{eqnarray}
\!\!\!\!\!\!\!\alpha_2(2\pi/L) = \frac{\alpha_2'}{1+ \frac{1}{\pi}\alpha_2'\left(1-12 {\alpha_2'}^2\sigma_2\frac{({\bf N}-2)}{L^2}\right)\ln{\bf N}\lambda} \ .
\label{eq:rglast}
\end{eqnarray}
The form of the extra contribution linear in ${\bf N}$ in the running
of the coupling $\alpha_2$ is entirely determined by the requirement
that the thermodynamic limit exist: i.e. that the ${\bf N}\ln {\bf N}$
in ${\cal K}(z)$ be cancelled by this extra piece.  Furthermore, we
know that this linear piece must scale as ${\bf N}-2$ since it is not
present in the two-body case. The necessity of this modification of the RG evolution
is of course no surprise as it is clear that the
thermodynamic limit of Eq.~(\ref{eq:rglast}) is equivalent to the
density-modified RG evolution of Eq.~(\ref{eq:rg1mod}).  Now, using
Eqs.~(\ref{eq:range3}) and (\ref{eq:range5mod}) and rescaling the
coupling using Eq.~(\ref{eq:rglast}), gives a finite result in the
thermodynamic limit
\begin{eqnarray}
\!\!\!\!\!\!\!{\cal E}^{\sigma_2}& =&\lim_{{\bf N}\to\infty} \rho\times \frac{E^{\sigma_2}_0}{{\bf N}}= 
-\frac{24{\alpha_2'}^{\!4} \rho^3\sigma_2}{\pi M }\bigg\lbrack \frac{7}{6}  + \ln\frac{\alpha_2'}{\lambda \pi^2}\bigg\rbrack,
\label{eq:energysig2B}
\end{eqnarray}
in perfect agreement with Eq.~(\ref{eq:energysig2}).

{\bf Conclusions.}  
Nonuniversal effects due to a non-vanishing effective range have been
computed for the weakly-coupled repulsive Bose gas in two spatial
dimensions using two distinct methods, both of which originate in the
most general EFT which describes bosons interacting at low-energies
via finite-range forces.  The first method is a perturbative expansion
about a mean field which gives directly the weak-coupling equation of
state. The fundamental assumption underlying this method is the
presence of the mean field; i.e. that the bosonic field $\psi$
acquires a vacuum expectation value $\sqrt{\rho}$. The second method
does not assume a mean field but rather computes the energy of a
finite number of bosons in a finite area and then takes the
thermodynamic limit. This latter, exact, method also gives an energy
that is perturbative in the coupling constant, but it is highly
singular in the number of bosons. The most singular terms in the
series are readily summed to give a result consistent with mean-field
theory.  The fundamental assumption underlying this method is the
existence of the thermodynamic limit. Like the three-dimensional case,
the ER corrections are highly suppressed as, in addition to the
momentum suppression, they necessarily arise from a loop effect. It
would be interesting to verify these new nonuniversal corrections
numerically using Monte Carlo simulations.

{\bf Acknowledgments.} 
I would like to thank Martin J.~Savage for a useful comment on the manuscript.
This work was supported in part by the U.~S.~Department of Energy
grant DE-SC001347.

\bibliographystyle{apsrev4-1}
\bibliography{bibi} 



\end{document}